\documentstyle[12pt,aasms4]{article}

\newcommand{\be}{\begin{equation}}
\newcommand{\ee}{\end{equation}}

\lefthead{Wei \& Lu}
\righthead{Can the jet steepen the light curve}

\begin{document}

\title{Can the Jet steepen the light curves of GRB afterglows?}

\author{D.M. Wei$^{1,2}$}
\affil{$^{1}$ Purple Mountain Observatory, Chinese Academy of Sciences, Nanjing, 210008, China\\
$^{2}$ National Astronomical Observatories, Chinese Academy of Sciences, China}
\and
\author{T. Lu$^{2,1,3}$}
\affil{$^{1}$ Department of Astronomy, Nanjing University, Nanjing, 210093, China\\
   $^{2}$ CCAST (World Laboratory), P.O. Box. 8730, Beijing, 100080, China\\
   $^{3}$ LCRHEA, IHEP, CAS, Beijing, China\\}

\begin{abstract}
Beaming of relativistic ejecta in GRBs has been postulated by many authors in order to reduce the total 
GRB energy, thus it is very important to look for the observational evidence of beaming. Rhoads (1999) has pointed 
out that the dynamics of the blast wave, which is formed when the beamed ejecta sweeping the external medium,
will be significantly modified by the sideways expansion due to the increased swept up matter. He claimed 
that shortly after the bulk Lorentz factor ($\Gamma $) of the blast wave drops below the inverse of the initial 
opening angle ($\theta _{0}$) of the beamed ejecta, there will be a sharp break in the afterglow light curves. 
However, some other authors have performed numerical calculations and shown that the break of the light curve
is weaker and much smoother than the one analytically predicted. In this paper we reanalyse the dynamical 
evolution of the jet blast wave, calculate the jet emission analytically,  we find that the sharp break 
predicted by Rhoads will actually not exist, and for most cases the afterglow light curve will almost
not be affected by sideways expansion unless the beaming angle is extremely small. We demonstrate
that only when $\theta _{0}<0.1$, the afterglow light curves may be steepened by sideways expansion,
and in fact there cannot be two breaks as claimed before. We have also constructed a simple numerical 
code to verify our conclusion. 
\end{abstract}


\section{Introduction}
The BeppoSAX results have revolutionized our understanding of GRBs by opening a window on X-ray,
optical and radio afterglows (e.g. Costa et al. 1997; Piro et al. 1998; Van Paradijs et al. 1997). The decaying
power-law long-wavelength afterglows are explained as the emission from a relativistic blast wave which
decelerates when sweeping up interstellar medium. The dynamical evolution of GRB fireballs and the emission
features have been studied by many authors (e.g. Sari 1997; Meszaros, Rees \& Wijers 1998; Wei \& Lu
1998a,b; Sari, Piran \& Narayan 1998), most of them considered the fireball being isotropic.

The discovery of GRB afterglow shows that GRBs are at cosmological distances. If so, the total energy 
for typical GRB event is about $10^{52}$ ergs. This year, an extraordinary event GRB990123 was
detected, which was the brightest burst ever detected by BeppoSAX satellite, and is in the top $0.3\%$ of
all bursts (Feroci et al. 1999). The detection of the redshift showed that the burst appears at $z\geq 1.6$,
with its $\gamma$-ray fluence of $\sim 5\times 10^{-4}$ ergs cm$^{-2}$, the total energy of this source is
$\geq 1.6\times 10^{54}$ ergs if the emission is isotropic (Andersen et al. 1999; Kulkarni et al. 1999).
This energy is so large that it gives a great challenge to the popular models. For models involving
stellar mass central engines it is necessary to assume that the ejecta are beamed in order to explain
such a huge energy. 

Now the main uncertainty of bursts' energy is whether the bursts radiate isotropically or are beamed into
a small solid angle. As shown above, the extreme large energy favors the emission being beaming. Then there
is one question: how can we identify the radiation between beaming or not? Rhoads (1997, 1999) have shown
that the lateral expansion of the shocked, relativistic plasma causes that at some moment the surface 
of the blast wave starts to increase faster than due to the cone-outflow alone, then the blast wave begins to 
decelerate faster than without the sideways expansion since more interstellar medium has been swept up 
by blast wave. Rhoads claimed that this effect will produce a sharp break in the GRB afterglow light curves.
Such a break is claimed to be present in the light curves of GRB990123 and GRB990510 (Kulkarni et al.
1999; Harrison et al. 1999). Sari et al. (1999) speculate that afterglows with very steep light curves are 
highly beamed.

However, some other authors (Panaitescu \& Meszaros 1998; Moderski et al. 1999) have performed numerical
calculations of dynamical evolution of blast wave, and shown that the break of the light curve is weaker and 
much smoother than the one analytically predicted. Thus there are two opposite conclusions about the jet
emission, the analytical treatment predicts a sharp break, while the numerical calculation shows no such 
sharp break.

In this paper we will first give an analytical treatment of the dynamical evolution of the jet blast wave and its
emission features, and will demonstrate that the sharp break will not actually exist, we may observe
the steepening of the light curve only when the jet angle is extremely small, i.e. $\theta _{0}<0.1$. We
then perform a simple numerical calculation to confirm our results. In next section 
we discuss the dynamical evolution of blast wave, in section 3 we 
calculate the jet emission analytically, and finally we give some discussions and conclusions.

\section{Dynamical evolution of the jet}

Now we consider an adiabatic relativistic jet expanding in surrounding medium. For energy conservation,
the evolution equation is 
\be
\Gamma ^{2}V={\rm const}
\ee
where $\Gamma $ is the bulk Lorentz factor, and $V$ is the jet volume, $V=2\pi r^{3}(1-cos\theta _{j})/3\propto 
r^{3}\theta _{j}^{2}$ for $\theta _{j}\ll 1$, and $\theta _{j}=\theta _{0}+\theta'=\theta _{0}+c_{s}t_{co}/ct$, where
$\theta _{0}$ is the initial jet opening half-angle, $\theta '$ describes the lateral expansion, $c_{s}$ is the
expanding velocity of ejecta material in its comoving frame, and $t\,(t_{co})$ is the time measured in the burster
frame (comoving frame). For  relativistic expanding material it is appropriate to take $c_{s}$ to be the sound
speed $c_{s}=c/3^{1/2}$ (Rhoads 1997, 1999), and Rhoads (1999) has given $t_{co}/t=2/5\Gamma $. Since the jet 
expands relativistically, there is the relation $T\propto r/\Gamma ^{2}$, where $T$ is the time measured in the 
observer frame, $r=ct$ is the radial coordinate in the burster frame. From above relations, we have
\be
\Gamma (1+\frac{\Gamma _{b}}{\Gamma })^{1/4}\propto T^{-3/8}
\ee
where $\Gamma _{b}=\frac{2}{5}\frac{c_{s}}{c}\theta _{0}^{-1}$. In this paper all quantities with the subscript "b"
denotes the point at which $c_{s}t_{co}=r\theta _{0}$, which means that after that time the sideways expansion
begins to dominate the radial divergence. Since $T\propto r/\Gamma ^{2}$, so $\Gamma _{b}/\Gamma =
(\frac{r_{b}}{r}\frac{T}{T_{b}})^{1/2}$. For the case $T\ll T_{b}$, it is well known that $\Gamma \propto T^{-3/8}$,
then $r\propto \Gamma ^{2}T\propto T^{1/4}$, and $\Gamma _{b}/\Gamma =(T/T_{b})^{3/8}$. For another case
$T\gg T_{b}$, the radial coordinate $r$ is nearly a constant, $r\approx r_{b}$, so $\Gamma _{b}/\Gamma \simeq 
(T/T_{b})^{1/2}$. Therefore we have
\be
\Gamma \propto \left \{
      \begin{array}{cc}
          T^{-3/8}[1+(\frac{T}{T_{b}})^{3/8}]^{-1/4}, & {\rm if}\,\, T<T_{b}\\
          T^{-3/8}[1+(\frac{T}{T_{b}})^{1/2}]^{-1/4}, & {\rm if}\,\, T >T_{b}
      \end{array}
  \right .
\ee
It is obviously that $\Gamma \propto T^{-3/8}$ for $T\ll T_{b}$, and $\Gamma \propto T^{-1/2}$ for $T\gg T_{b}$.
The rapid decrease with time of Lorentz factor $\Gamma $ is due to the fact that larger amounts of surrounding
matter has been swept up by ejecta (Rhoads 1997, 1999).

In the following we calculate the value of $T_{b}$. According to the fireball model, the decelerating radius of 
the ejecta is
\be
r_{d}=(\frac{E}{\pi \theta _{0}^{2}\Gamma _{0}^{2}n_{1}m_{p}c^{2}})^{1/3}
\ee
where $E$ is the burst energy, and $\Gamma _{0}=E/M_{0}c^{2}$, $M_{0}$ is the initial baryon mass. 
Rhoads (1999) has given $r_{b}=[\frac{75\Gamma _{0}^{2}\theta _{0}^{2}}{8(c_{s}/c)^{2}}]^{1/3}r_{d}$. 
The relation between $T$ and $r$ is $T=r/\zeta \Gamma ^{2}c$, where the numerical value of $\zeta $ 
lies between $\sim 3$ and $\sim 7$ depending on the details of the hydrodynamical evolution and the
spectrum. Sari (1997) and Waxman (1997) have shown that the typical value of $\zeta $ is about 4, then 
the time
\be
T_{b}\simeq \frac{r_{b}}{4\Gamma _{b}^{2}c}=70(\frac{c_{s}}{c/\sqrt{3}})^{-8/3}(\frac{\theta _{0}}{0.1})^{2}E_{52}^{1/3}
n_{1}^{-1/3}\,\,\,(day)
\ee
We see that the break time $T_{b}$ is very large for typical parameters, which means that the transition 
from $\Gamma \propto T^{-3/8}$ to $\Gamma \propto T^{-1/2}$ is usually very slowly and smoothly.

\section{The emission from jet}

Now we calculate the emission flux from the jet. Here we adopt the formulation and notations of Mao \&
YI (1994). In our model the ejecta is flowing outwards relativistically  (with Lorentz factor $\Gamma $) in 
a cone with opening half angle $\theta _{j}$. For simplicity, we assume that the radiation is isotropic
in the comoving frame of the ejecta and has no dependence on the angular positions within the cone.
The radiation cone is uniquely defined by the angular spherical coordinates ($\theta $,$\phi $) of its
symmetry axis, here $\theta $ is the angle between the line of sight (along $z$-axis) and the symmetry
axis, and $\phi $ is the azimuthal angle. Because of cylindrical symmetry, we can assume that the 
symmetry axis of the cone is in the $y-z$ plane. In order to see more clearly, let us establish an
auxiliary coordinate system ($x', y', z'$) with the $z'$-axis along the symmetry axis of the cone and the
$x'$ parallel the $x$-axis. Then the position within the cone is specified by its angular spherical 
coordinates $\theta '$ and $\phi '$ ($0\leq \theta '\leq \theta _{j}$, $0\leq \phi '\leq 2\pi $). It can be shown that
the angle $\Theta $ between a direction ($\theta ',\,\phi '$) within the cone, and the line of sight satisfies
$cos\Theta =cos\theta cos\theta '-sin\theta sin\theta 'sin\phi '$. Then the observed flux is
\be
F(\nu ,\theta )=\int_{0}^{2\pi }d\phi '\int_{0}^{\theta _{j}}sin\theta 'd\theta' D ^{3}
I'(\nu D^{-1})\frac{r^{2}}{d^{2}}
\ee
where $D=[\Gamma (1-\beta cos\Theta )]^{-1}$ is the Doppler factor, , $\beta =(1-\Gamma ^{-2})^{1/2}$, 
 $\nu =D\nu '$, $I'(\nu ')$ is the specific intensity of synchrotron radiation at $\nu '$, and $d$ is the 
distance of the burst source. Here the quantities with prime are measured in the comoving frame. For
simplicity we have ignored the relative time delay of radiation from different parts of the cone.

For the expanding jet, we have $r=D\Gamma \beta cT\propto D\Gamma T\,(\beta \simeq 1)$, $r'=D\beta cT
\propto DT$, the magnetic field strength $B'\propto \Gamma $, the peak frequency of synchrotron radiation
$\nu _{m}=D\nu _{m}'\propto D\Gamma ^{3}$, and $I'(\nu _{m}')\propto n_{e}'B'r'
\propto D\Gamma ^{2}T$. Assuming that the emission spectrum $I'(\nu ')\propto \nu '^{-\alpha }$, then
$I'(\nu ')=I'(\nu _{m}')(\frac{\nu '}{\nu _{m}'})^{-\alpha }=I'(\nu _{m}')(\frac{\nu }{\nu _{m}})^{-\alpha }
\propto D^{1+\alpha }\Gamma ^{2+3\alpha }T\nu ^{-\alpha }$. Therefore we have the flux 
\be
F(\nu ,\theta )\propto \nu ^{-\alpha }\Gamma ^{2(\alpha -1)}T^{3}g(\theta ,\Gamma ,\alpha )
\ee
where
\be
g(\theta ,\Gamma ,\alpha )=\int_{0}^{2\pi }d\phi '\int_{0}^{\theta _{j}}sin\theta 'd\theta' 
(1-\beta cos\Theta )^{-(6+\alpha )}
\ee
In general, the value of $g$ can only be calculated numerically. However here we consider the case
$\theta _{j}\ll 1$ and $\theta \ll 1$, then $cos\Theta \approx cos\theta cos\theta '$. In this case we
can calculate the value of $g$ analytically under certain conditions. After complicated calculation
we find $g\propto \theta ^{-2(5+\alpha )}$ for $\Gamma ^{-1}<\theta <\theta _{j}$, $g\propto 
\theta _{j}^{2}\theta ^{-2(5+\alpha )-2}$ for $\Gamma ^{-1}<\theta $ and $\theta >\theta _{j}$, $g\propto 
\Gamma ^{2(5+\alpha )}$ for $\theta <\Gamma ^{-1}<\theta _{j}$, and $g\propto \Gamma ^{2(5+\alpha )+2}
\theta _{j}^{2}$ for $\theta <\Gamma ^{-1}$ and $\theta _{j}<\Gamma ^{-1}$. Therefore we have the results
\be
F(\nu ,\theta )\propto \left \{
    \begin{array}{ll}
      \nu ^{-\alpha }\theta ^{-2(5+\alpha )}\Gamma ^{2(\alpha -1)}T^{3}, & {\rm for}\,\,\Gamma ^{-1}<\theta<\theta_{j}\\   
      \nu^{-\alpha}\theta_{j}^{2}\theta^{-2(5+\alpha)-2}\Gamma^{2(\alpha-1)}T^{3},& {\rm for} \,\,\Gamma^{-1}<\theta ,\,\,
                          \theta >\theta _{j}\\
      \nu ^{-\alpha }\Gamma ^{8+4\alpha }T^{3}, & {\rm for}\,\,\theta <\Gamma ^{-1}<\theta _{j}\\
      \nu ^{-\alpha }\theta _{j}^{2}\Gamma ^{10+4\alpha }T^{3}, & {\rm for}\,\,\theta <\Gamma ^{-1},\,\theta _{j}<\Gamma^{-1}
    \end{array}
   \right .
\ee
For $T\ll T_{b}$, the evolution is about $\Gamma =\Gamma _{0}(r/r_{d})^{-3/2}$. Here we define the
Lorentz factor, $\Gamma _{j}\equiv \Gamma (r_{j})=\theta _{j}^{-1}$, then $r_{j}=(\Gamma _{0}\theta _{j})^{2/3}r_{d}$, 
and the corresponding timescale $T_{j}\simeq r_{j}/4\Gamma _{j}^{2}c
\simeq 1.3(\frac{\theta _{j}}{0.1})^{2}E_{52}^{1/3}n_{1}^{-1/3}(\frac{\theta _{j}}{\theta _{0}})^{2/3}$ day. Then 
the observed flux at fixed frequency is
\be
F\propto \left \{
   \begin{array}{ll}
      T^{-3\alpha /2}, & {\rm for}\,\,T<T_{j}\\
      T^{-\frac{3}{2}\alpha -\frac{3}{4}}, & {\rm for}\,\,T_{j}<T<T_{b}\\
      T^{-2\alpha -1}, & {\rm for} \,\,T_{b}<T
    \end{array}
   \right .
\ee
From above it seems that there should be two temporal index breaks in light curves. However,
if we compare the values of $T_{j}$ and $T_{b}$, we will find that $T_{j}\ll T_{b}$, i.e. the time interval
between $T_{j}$ and $T_{b}$ is very large, the beaming break is much earlier than the break due to 
sideways expansion. We know that in order to see the steepening of the light curve, $T_{j}$ or 
$T_{b}$ must be small, so in fact we can only see one temporal break. In addition, in the Rhoads'
treatment, the effect of the sideways expansion on the $\Gamma $ evolution was ignored when
$T<T_{b}$, however in fact, there is still some sideways expansion during this phase, so the 
evolution of $\Gamma$ must be affected by sideways expansion when $\Gamma \sim \theta _{0}^{-1}$.
Therefore, we expect that the evolution of $\Gamma$ is continuous, and the transition from
$\Gamma \propto T^{-3/8}$ to $\Gamma \propto T^{-1/2}$ is much smoother than previously claimed.

In order to test our conclusion, we make a simple numerical calculation. We assume that the blast
wave evolution is adiabatic, ignore cooling of the swept-up particles. We take the following initial
parameters: $\Gamma _{0}=300$, the electron distribution index $p=2.5$. In Fig.1 we present the
light curve for different initial opening angle, the solid, dotted and dashed lines represent the 
cases where $\theta _{0}=0.1,\,\,0.0174,\,\,0.01$ respectively. We show that when $\theta _{0}=0.1$,
the light curve is nearly not affected by sideways expansion, while when $\theta _{0}=0.0174$ and 
0.01, the light curves are steepened clearly, confirming our analytic conclusion.

As for comparison, we also calculate the afterglow light curves of blast wave with no spreading.
We know that, if the blast wave is highly radiative, the internal energy of the blast wave will be
low, since it is converted to photons and radiated away, so the lateral expanding velocity will be
very small, $c_{s}<<c$, in this case, the sideways expansion is unimportantly small, and the blast
wave can be regarded as no spreading. For highly radiative evolution, it has been shown that the Lorentz
factor $\Gamma \propto T^{-3/7}$ (Wei \& Lu 1998a). Taking the parameters as above, we calculate
the afterglow light curves under this situation, Fig.2 gives our results, the solid, dotted and dashed 
lines also correspond to the cases $\theta _{0}=0.1,\,\,0.0174,\,\,0.01$ respectively. It is obvious that,
when $T<T_{c}$ ($T_{c}\sim T_{0}(\Gamma _{0}\theta _{0})^{7/3}$), the light curves decay as a simple
power law, and when $T>T_{c}$, the light curves deviate from the simple power law and show a 
steepening, and a clear break occurs at about $T_{c}$. Also we note that only when $\theta _{0}<0.1$,
the steepening is obvious. Therefore we suggest that if we observe the sharp break in the GRB
afterglow light curves, then it may indicate that the blast wave is highly radiative rather than adiabatic.

\section{Discussion and conclusions}

The GRB afterglows provide very good opportunity to study whether and how much the GRB ejecta are
beamed. Rhoads (1997, 1999) has pointed out that the beamed outflows should diverge from the cone
geometry and the sideways outflow of the shocked relativistic plasma would increase the front of the
blast wave which leading to a fast deceleration. He also predicted that the afterglow light curves should
have a sharp break around $T_{b}$.

However, Moderski et al. (1999) have performed numerical calculation and shown that the break of the
light curve is weaker and smoother than the prediction. Here we reanalyse the dynamical evolution 
of the jet blast wave, calculate the emission from the jet. Our calculations show that the main reason 
why the results of Moderski et al.  being different
from that of Rhoads is that the value of $T_{b}$ is very large when taking the parameters adopted 
by Modersli et al. (see eq. 5). Our formular (eq. 3) indicates that the evolution of Lorentz factor $\Gamma $
with time $T$ is continuous, changing the slope from -3/8 to -1/2 smoothly. In particular, if the value of 
$T_{b}$ is large, then the transition is much smoothly, in this case one expects that the sharp break
will not exist. In order to test our analytic conclusion, we also make a simple numerical calculation,
from Fig.1 it is shown that only when $\theta _{0}<0.1$, we can observe the steepening of the 
light curve, which is consistent with our analytic conclusion.

Our results are valid only if the remnant is still relativistic at time $T_{b}$. Since at this time the Lorentz
factor $\Gamma _{b}\simeq \frac{2}{5}\frac{c_{s}}{c}\theta _{0}^{-1}$, this condition reduces to $\theta _{0}
<0.1$, i.e. the jet is very narrow. If $\theta _{0}>0.1$, then before the sideways expansion is important, 
the remnant has already become non-relativistic. In fact, Dai and Lu (1999) have shown that even in
the case of isotropic emission, the break in the light curve can appear during the transition from
ultrarelativistic to non-relativistic phase in the environment of dense material. 

It should be emphasized that in our calculation we have ignored the relative time delay of radiation
from different parts of the cone, if this effect is considered, then the slope of the light curve should be
flatter than $T^{-(2\alpha +1)}$ (Moderski et al. 1999). However, all these calculation are based on the 
assumption that the material is uniformly distributed across the blast wave. In fact, it is more likely 
that the lateral outflow can lead to $\theta $ dependent structure, with the density of swept material and
the bulk Lorentz factor decreasing with $\theta $, in this case the break in the light curve may become
more prominent.

\acknowledgements{We thank the referee for several important comments that improved this paper.
This work is supported by the National Natural Science Foundation (19703003 and 
19773007) and the National Climbing Project on Fundamental Researches of China.}

\newpage

\vspace{25mm}
\centerline{\Large Figure Caption}

\vspace{25mm}

Fig.1 The afterglow light curves for  different initial opening angle, the solid, dotted and dashed lines
represent the cases where $\theta_ {0}=0.1,\,\,0.0174$ and 0.01 respectively.
 
Fig.2 The afterglow light curves for highly radiative blast wave with no spreading. The solid, dotted and 
dashed lines correspond to the cases where $\theta_ {0}=0.1,\,\,0.0174$ and 0.01 respectively.

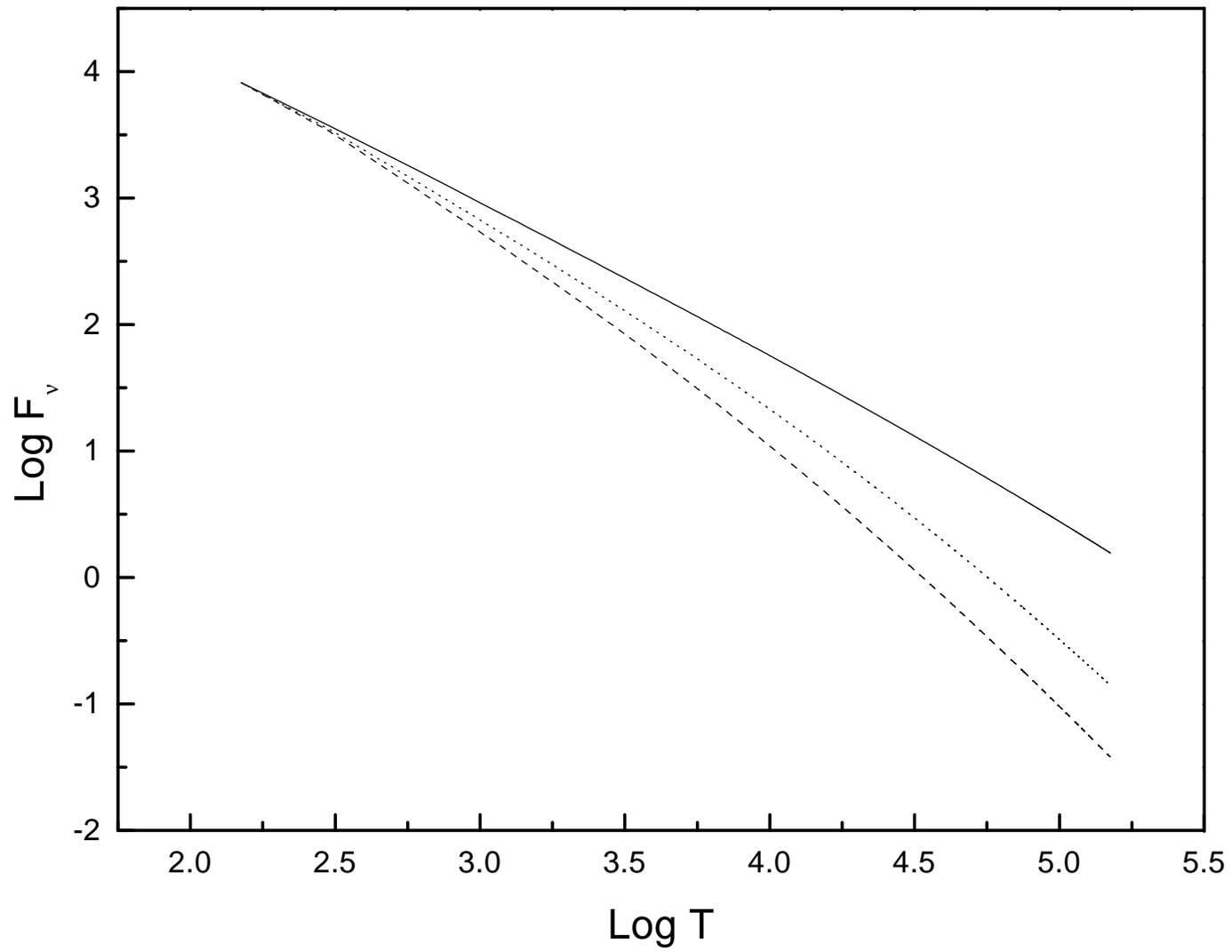
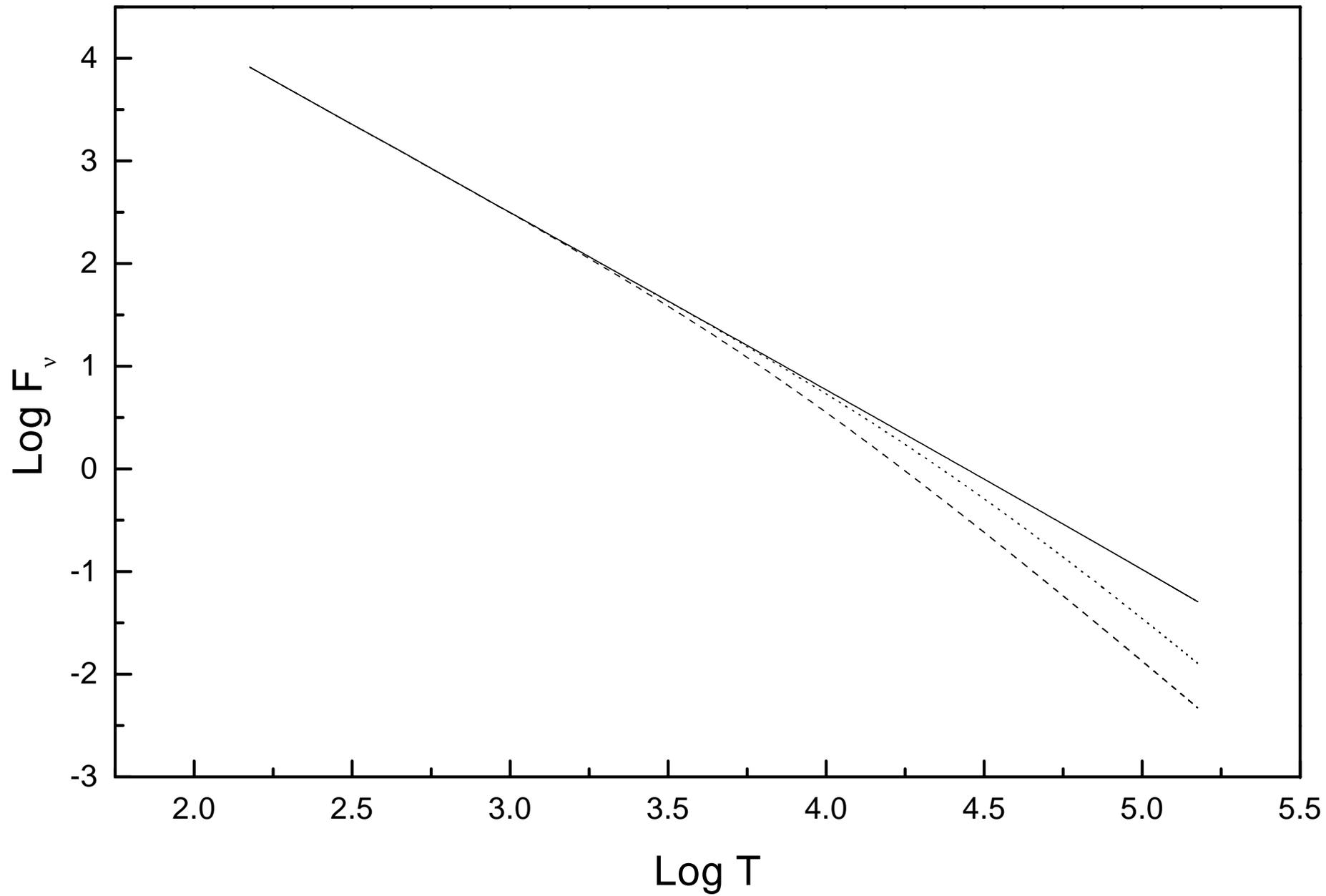
\end{document}